\documentclass[twocolumn,preprintnumbers,amsmath,amssymb]{revtex4}
\begin{document}
\title{Spherically symmetric curved space times from quantum fields backreaction corrections in two dimensional analogue}
\author{HOSSEIN GHAFFARNEJAD}
\affiliation{Physics Department, Semnan University, Semnan, IRAN,
 Zip code: 35131-19111}

\altaffiliation{Email address: hghafarnejad@yahoo.com}

 \maketitle
  \textbf{\textsl{ABSTRACT}-}
  Aim of the
paper is to obtain 2d analogue of the backreaction equation which
will be useful to study final state of quantum perturbed spherically
symmetric curved space times. Thus we take Einstein-massless-scalar
$\psi$ tensor gravity model described on class of spherically
symmetric curved space times. We rewrite the action functional in 2d
analogue in terms of dimensionless dilaton-matter field
$(\chi=\Phi\psi)$ where dilaton field $\Phi$ is conformal factor of
2-sphere. Then we seek renormalized expectation value of quantum
dilaton-matter field stress tensor operator by applying Hadamard
rennormalization prescription. Singularity of the Green function is
assumed  to be has logarithmic form. Covariantly conservation
condition on the renormalized quantum dilaton-matter stress tensor
demands to input a variable  cosmological parameter $\lambda(x)$.
Energy conditions (weak, strong and null) is studied on the obtained
renormalized stress tensor leading to dynamical equations for
$\lambda(x), \Phi$ and quantum vacuum state
$W_0(x)=<0|\hat{\chi}^2|0>_{ren}.$ In weak quantum field limits our
obtained trace anomaly corresponds to one which obtained from zeta
function regularization method. Setting null-like apparent horizon
equation $\nabla_c\Phi\nabla^c\Phi=0,$ our procedure predicts that
physically correct value of the parameter in the anomaly trace
$\frac{1}{24\pi}\{R-\alpha\frac{\nabla_{c}\nabla^{c}\Phi}{\Phi}+(\alpha-6)\frac{\nabla_c\Phi\nabla^c\Phi}{\Phi^2}
\}$ should be $\alpha=6.$ At last we solved the backreaction
equation and obtained explicitly metric field solution in the slow
varying limits of the quantum and dilaton fields which has black
holes topology and its singularity is covered by apparent horizon
hypersurface. \vskip 0.50 cm
 \textbf{\textsl{Keywords}}-Dilaton fields, Dimensional reduction,
 Hadamard renormalization, Spherically symmetric curved space times,
 Variable cosmological parameter
\section{\label{I}Introduction}
In absence of a viable theory of pure quantum gravity, its
semiclassical approximation readily yields particle creation in
curved background space time (see [1,2] and references therein). In
the latter approach the gravitational field is retained as a
classical background, while the matter fields are quantized in the
usual way. In the latter view the perturbed metric is obtained by
the semiclassical Einstein beakreaction equations.
\begin{equation} G_{\mu\nu}=8\pi
G\{T^{class}_{\mu\nu}+<\hat{T}_{\mu\nu}>_{ren}\}\end{equation} where
we used units $c=\hbar=1,$ $G_{\mu\nu}\equiv
R_{\mu\nu}-\frac{1}{2}g_{\mu\nu}R$ with $\mu,\nu=0,1,2,3$ is
Einstein tensor in four dimensional curved space-time, $R_{\mu\nu}$
(R) is Ricci tensor (scalar). $T^{class}_{\mu\nu}$ is classical
matter fields stress tensor. $<\hat{T}_{\mu\nu}>_{ren}$ is
renormaized expectation value of quantum matter fields operator.
According to wald`s axioms [3], $<\hat{T}_{\mu\nu}>_{ren}$ must be
covariantly conserved $\nabla^\mu<\hat{T}_{\mu\nu}>_{ren}=0,$ but in
the presence of trace anomaly. For conformaly invariant fields the
trace anomaly $<\hat{T}^\nu_\nu>_{ren}$ is nonzero, unlike its
classical counterpart, and is independent of the quantum state where
the expectation value is taken. It is completely expressed in terms
of geometrical objects as
\begin{equation}<\hat{T}^\nu_{\nu}>_{ren}=\frac{1}{2880\pi^2}\{aC_{\alpha\beta
\gamma\delta}C^{\alpha\beta\gamma\delta}+b(R_{\alpha\beta}R^{\alpha\beta}-R^2/3)$$$$+
c\nabla_{\gamma}\nabla^{\gamma} R+dR^2\}\end{equation} where
$a,b,c,d$ are known as depended on the spin of the quantum fields
under consideration [1,2] and $C_{\alpha\beta\gamma\delta}$ is Weyl
tensor. Whether such an approach makes sense is subject to debate.
Due to the non-linearity of gravity, it will certainly fail for
effects that occur on the scale of the Planck length
$(G\hbar/c^3)^{1/2}=1.616\times10^{-33} cm,$ or involve
singularities. Thus it will certainly not be possible to correctly
describe, among other things, the very final stage of black holes
evaporation in a semiclassical model. On the other hand, one might
expect meaningful results as long as one stays in the region
exterior of a reasonably sized black hole. It is hopped that the
semiclassical approximation in gravity works similarly to the
quantum electrodynamics one which is able to describe quantum
particles in exterior
electromagnetic fields. \\
Yet in the semiclassical approximation as well as in full quantum
gravity, the equations describing the evolution of the system must
be solved self-consistently. In four dimensions, this poses a
problem: One is only able to calculate the Hawking radiation for a
fixed spherically symmetric background metric. Even in the latter
case, one obtain instead a relation constraining undetermined
function [4] and so study of black hole Hawking radiation in four
dimension exhibits with some little success. Hence the Hawking
radiation and backreaction effects of created particles on the
dynamical background metric is still as an open
problem.\\
In order to get a suitable answer to this problem, one takes
two-dimensional analogue of the gravitational models from four
dimensions by introducing a dilaton field which contains physical
properties of tangential pressure of (classical and quantum) matter
fields. The latter idea is a good proposal and zeta function
regularization method is used to obtain effective action functionals
and corresponding anomaly trace in literature [5,6,7,8]. In this
paper we use other procedure called with Hadamard renormalization
prescription . Our procedure inputs a variable cosmological
parameter $\lambda(x)$ reaching to the covariantely conservation
condition of the renormalized stress tensor. This variable
cosmological
 parameter is really corrections of an essential effective cosmological constant $\Lambda_{eff}=\frac{1}{4\pi G}$
defined by Newtonian coupling constant $G$ which comes from
dimensional reduction of space time [9] (see also [10,11,12]
for other applications of variable cosmological parameter idea).\\
 Organization of the paper
is as follows.
 In section II we review
two dimensional analogue of the Einstein-Hilbert gravity minimally
coupled with mass-less scalar field propagating in s-mode.
 In section III we suggest
symmetric two-point Hadamard Green function to be contained
logarithmic geometrical singularity. This suggestion is originated
from corresponding to Green function of a massless scalar field
moving on two dimensional Minkowski flat space time [13]. Hadamard
renormilization prescription makes ultraviolet singularities of all
physical objects such as, quantum matter action functional, stress
tensor expectation value of the field and etc., same as logarithmic
geometrical singularity. Renormalized expectation value of the
quantum matter stress tensor operator leads to a nonsingular
covariantly conserved stress tensor with  anomaly trace in the
presence of variable cosmological parameter. The suggested variable
cosmological parameter is described in terms of derivatives of the
dilaton field, Ricci scalar of induced 2d
 background metric and derivatives of quantum vacuum state $W_0(x)$.
  In section IV we study energy conditions (weak, strong, null) on the obtained renormalized stress tensor
  which leads to dynamical equations
  of the fields $\Phi,W_0(x),\lambda(x).$ Also our procedure in weak quantum field limits
  follows results of the one which obtained
  from zeta function regularization method. In section V we use apparent horizon property
  of the curved space times on the obtained
  anomaly trace of quantum matter field stress tensor expectation value.
  Section VI denotes to concluding remarks.
\section{\label{II}The Model}
We take Einstein-Hilbert gravity interacting with massless scalar
matter field $\psi$ in 4d curved space times \begin{equation}
I=\frac{1}{16\pi G}\int
dx^4\sqrt{\tilde{g}}\tilde{R}-\frac{1}{2}\int
dx^4\sqrt{\tilde{g}}\tilde{g}^{\mu\nu}\partial_\mu\psi\partial_{\nu}\psi\end{equation}
where $\tilde{g}$ is absolute value of determinant of the 4d curved
space time metric $\tilde{g}_{\mu\nu}$ $(\mu,\nu=0,1,2,3)$ and
$\tilde{R}$ is its Ricci scalar. Varying the above action with
respect to the fields $\tilde{g}^{\mu\nu}$ and $\psi$ one can obtain
corresponding field equations as \begin{equation}
\tilde{G}_{\mu\nu}=8\pi G \tilde{T}_{\mu\nu}\end{equation} and
\begin{equation}\tilde{\nabla_{\gamma}}\tilde{\nabla^{\gamma}}\psi=0\end{equation} where  \begin{equation}
\tilde{T}_{\mu\nu}=\partial_{\mu}\psi\partial_{\nu}\psi-
\frac{1}{2}\tilde{g}_{\mu\nu}\{\partial_{\gamma}\psi\partial^{\gamma}\psi\}\end{equation}
in which Bianchi identity $\tilde{\nabla}^{\mu}\tilde{G}_{\mu\nu}=0$
leads to covariant conservation condition of the matter field
\begin{equation} \tilde{\nabla}^{\mu}\tilde{T}_{\mu\nu}=0,~~~
\tilde{T}_{\mu}^{\mu}=-\tilde{g}^{\mu\nu}\partial_{\mu}\psi\partial_{\nu}\psi.\end{equation}
We choose class of 4d spherically symmetric curved space times
metrics as \begin{equation}
ds^2=\tilde{g}_{\mu\nu}dx^{\mu}dx^{\nu}$$$$=g_{ab}(x^a)dx^adx^b+
\Phi^2(x^a)(d\theta^2+\sin^2\theta d\phi^2)\end{equation} where
signature of the metric (8) is assumed to be $(-,+,+,+).$ Then we
assume that the metric fields $g_{ab},$ $\Phi$ and matter field
$\psi$ are independent of angular coordinates $(\theta,\phi)$
propagating in spherically modes (S-channel) and integrate (3) with
respect to angular coordinates $\theta$ and $\phi$ leading to [9]
\begin{equation} I=\frac{1}{4G}\int
dx^2\sqrt{g}\{1+g^{ab}\partial_a\Phi\partial_b\Phi+\frac{1}{2}\Phi^2R\}$$$$-2\pi\int
dx^2\sqrt{g}\Phi^2\partial_a\psi\partial^a\psi.\end{equation} $\Phi$
is called geometrical dilaton  field with $length$ dimensions and it
is in agreement with the status of boson particles in point of view
of field theory. $g_{ab}(x^0,x^1)$ is 2d induced metric on the
hypersurface $\theta=\phi=constant$.
 $g$ is absolute value of determinant of 2d metric
$g_{ab}$ and $R$ is corresponding 2d Ricci scalar. The matter field
$\psi$ has inverse of length dimensions. Varying (9), with respect
to $g_{ab},$ $\Phi,$ and $\psi,$ the corresponding field equations
are obtained respectively as [9] \begin{equation}\Phi^2
\tilde{G}_{ab}=-2\Phi\nabla_a\nabla_b\Phi+g_{ab}\{2\Phi\nabla_{c}\nabla^{c}\Phi+\partial_c\Phi\partial^c\Phi-1\}$$$$=8\pi
G\Phi^2\tilde{T}_{ab}[\psi],\end{equation} \begin{equation}
\tilde{G}_{\theta\theta}=\Phi\nabla_{c}\nabla^{c}\Phi-\frac{1}{2}R\Phi^2=8\pi
G\tilde{T}_{\theta\theta}=-4\pi
G\Phi^2\partial_c\psi\partial^c\psi,\end{equation} and
\begin{equation}
\nabla_{c}\nabla^{c}\psi=-2J^a\nabla_a\psi,~~~J_a=\nabla_a\ln\Phi\end{equation}
where we defined  \begin{equation}
\nabla_a\nabla^a=\frac{1}{\sqrt{g}}\partial_a(\sqrt{g}g^{ab}\partial_b),~~~\partial_a\equiv\frac{\partial}{\partial
x^a} \end{equation} and non-angular components of the stress tensor
(6) as \begin{equation}
\tilde{T}_{ab}[\psi]=\partial_a\psi\partial_b\psi-\frac{1}{2}g_{ab}\partial_c\psi\partial^c\psi.\end{equation}
 The
 matter stress tensor (14) is trace
free but same as (7) dose not satisfy the covariant conservation
condition in 2d space times. Applying (13) and (14) one can obtain
\begin{equation}\nabla^aT_{ab}=-2J_a\partial^a\psi\partial_b\psi,~~~T^a_a=0\end{equation}
where we are dropped over tilde $^\sim.$ Violation of covariant
conservation is caused because of non-vanishing dilaton current
$J_a$ and it is coupled with matter current $\partial_a\psi$ as a
source in RHS of the matter wave equation (12). The quantity
$J_a\partial^a\psi$ treats as scalar charge for the field $\psi$
from view of string theory. Originally this charge comes from
dynamical effects of reference frames. For instance in higher
dimensional string theory of gravity the Brans-Dicke scalar tensor
theory is charge-less and so a covariantly conserved model in Jordan
frame but it is not in other frames

(see Ref. [14] chapter 2). Hence the string theory accepts that the
Bianchi identity no longer implies the covariant conservation of the
stress tensors separately in 4+d dimensional curved space times. In
other words stress tensors of matter and geometrical dilaton fields
do not need follow covariant conservation conditions separately.
Physically non-conservation condition of the stress tensor implies
that the motion of a free test  particle is no longer geodesic when
the particle has an intrinsic scalar charge and the gravitational
background contains a
non-trivial dilaton component.\\
However we follow here other point of view: dimensional reduction of
the space times causes to break covariant conservation condition.
  On the other hand we know that
renormalization of the quantum matter fields breaks also the
covariant conservation condition of the stress tensor (see [1,2] and
references therein). Some applicable methods are presented to
satisfy the covariant conservation condition but by inducing anomaly
trace. What is correspondence between them to obtain both quantum
matter stress tensor and its geometrical classical dilaton counter
part satisfying covariantly conservation condition separately in 2d
gravity model (9)? In the following section we try to obtain a
suitable answer to this question. We apply Hadamard renormalization
prescription to evaluate regular expectation value of quantum matter
stress tensor operator $<\hat{T}_{ab}[\hat{\psi}]>_{ren}$ by
presenting a variable cosmological parameter $\lambda(x).$
\section{Hadamard Renormalization}
If $\psi$ treats as massless quantum
  bosons. Then it will be linear operator $\hat{\psi}$
operating on arbitrary state of Hilbert space. Corresponding stress
energy tensor operator $\hat{T}_{ab}[\hat{\psi}]$ become bi-linear
with respect to $\hat{\psi}$ and regular stress tensor counterpart
${<\hat{T}_{ab}[\hat{\psi}]>}_{ren}$  (subscript $`ren`$ denotes to
$`renormalized`$) is obtained by eliminating its ultraviolet
divergence terms, in one loop level. With given
 ${<\hat{T}_{ab}[\hat{\psi}]>}_{ren}$ one can write two dimensional analogue of the
metric back reaction equation (1) by regarding (10) and (11) such as
follows. \begin{equation}
G_{ab}=-\frac{2\nabla_a\nabla_b\Phi}{\Phi}+g_{ab}\{\frac{2\nabla_{c}\nabla^{c}\Phi}{\Phi}+
\frac{\partial_c\Phi\partial^c\Phi}{\Phi^{2}}-\frac{1}{\Phi^{2}}\}$$$$=8\pi
G\frac{<\Phi^2\hat{T}_{ab}[\hat{\psi}]>_{ren}}{\Phi^{2}}\end{equation}
and \begin{equation}
G_{\theta\theta}=\Phi\nabla_{c}\nabla^{c}\Phi-\frac{1}{2}R\Phi^2
=-4\pi
G<\Phi^2\partial_c\hat{\psi}\partial^c\hat{\psi}>_{ren}\end{equation}
where $g_{ab}$ and $\Phi$ is still treats as classical geometrical
fields whereas the matter field $\psi$ is assumed to be treat as
quantum field. Furthermore we would not move $\Phi$ outside the
expectation quantities $<\Phi^2\hat{T}_{ab}[\hat{\psi}]>_{ren}$ and
$<\Phi^2\partial_c\hat{\psi}\partial^c\hat{\psi}>_{ren},$ because
variable dilaton field causes to violation of covariantly
conservation of matter stress tensor $T_{ab}[\psi]$ in its classical
regime (see Eq. (15)). Applying (16) and (17) the Bianchi identity
$\nabla^{\mu}G_{\mu\nu}=0$ in 4d leads to the following constraint
condition. \begin{equation}
\nabla^a<\Phi^2\hat{T}_{ab}[\hat{\psi}]>_{ren}=\nabla_b
\bigg(\frac{1}{\Phi^2}\bigg)<\Phi^2\partial_c\hat{\psi}\partial^c\hat{\psi}>_{ren}.\end{equation}
In 4d space times $\Phi$ and $1/\psi$ has length dimension and
conformal invariance property of the matter action in (3) is broken
in 2d analogue (9). Hence it will be useful we define a
dimensionless dilaton-matter field as \begin{equation}
\chi=\Phi\psi\end{equation} before than that we proceed to apply
renormalization prescription and evaluate expectation value of its
stress tensor operator $<\hat{T}_{ab}[\hat{\chi}]>_{ren}.$ Applying
(19), one can rewrite matter part of the action (9) as
\begin{equation} I_{matter}[\chi,g_{ab}, \Phi]=2\pi\int
\sqrt{g}dx^2g^{ab}\{\nabla_a\chi\nabla_b\chi+\chi^2J_aJ_b$$$$-\chi
J_b\nabla_a\chi-\chi J_a\nabla_b\chi\}.\end{equation}  Dynamical
equation of the field $\chi$ is obtained by varying the above action
with respect to $\chi$ as \begin{equation}
\{\nabla_{c}\nabla^{c}-\frac{\nabla_{c}\nabla^{c}\Phi}{\Phi}\}\chi=0\end{equation}
Trace free $T^a_a[\chi]$ stress tensor of the field $\chi$ is
obtained by varying (20) with respect to $g^{ab}$ such as follows.
\begin{equation}
T_{ab}[\chi]=\nabla_a\chi\nabla_b\chi+\chi^2J_aJ_b-\chi(J_a\nabla_b\chi+J_b\nabla_a\chi)$$$$
-\frac{g_{ab}}{2}\{\nabla_c\chi\nabla^c\chi+\chi^2J_cJ^c-2\chi
J_c\nabla^c\chi\}\end{equation} which is equivalent with
$\Phi^2T_{ab}[\psi]$
 and so we can deduce \begin{equation}
<\Phi^2\hat{T}_{ab}[\hat{\psi}]>\equiv<\hat{T}_{ab}[\hat{\chi}]>\end{equation}
and \begin{equation}
<\Phi^2\partial_c\hat{\psi}\partial^c\hat{\psi}>\equiv$$$$
<\partial^c\hat{\chi}\partial_c\hat{\chi}>-2J^c<\hat{\chi}\partial_c\hat{\chi}>+J_cJ^c<\hat{\chi}^2>.\end{equation}
In the following we seek renormalized expectation values of the
quantities (22) and (24) by applying the Hadamard
renormalization prescription.\\
 This approach is begun with definition
of the expectation value of stress tensor (22) such as follows.
\begin{equation} <\hat{T}_{ab}[\hat{\chi}]>=\lim_{x^{\prime}\to
x}D_{ab}(x,x^{\prime})G^+(x,x^{\prime}) \end{equation} where a state
of $\hat{\chi}$ is characterized by a hierarchy of Wightman function
which for a symmetric two-point function we have  \begin{equation}
G^+(x,x^{\prime})=\frac{1}{2}<\hat{\chi}(x)\hat{\chi}(x^{\prime})+\hat{\chi}(x^{\prime})\hat{\chi}(x)>\end{equation}
and \begin{equation}
D_{ab}(x,x^{\prime})=g_b^{b^{\prime}}\nabla_a\nabla_{b^{\prime}}+g_a^{a^{\prime}}\nabla_{a^{\prime}}
\nabla_b+J_aJ_b$$$$-J_a\{\nabla_b+g^{b^{\prime}}_b\nabla_{b^{\prime}}\}-
J_b\{\nabla_a+g^{a^{\prime}}_a\nabla_{a^{\prime}}\}$$$$
-g^{ab}\{g^c_{c^{\prime}}\nabla_c\nabla^{c^{\prime}}-J_c(\nabla^c+\nabla^{c^{\prime}})+\frac{J_cJ^c}{2}
\}\end{equation} with  the bivector of parallel transport
$g^a_{a^{\prime}},$ is the bilocal differential operator. This
expression makes explicit that the singular character of the
operator $\hat{T}_{ab}$ emerges as a consequence of the
short-distance singularity of the symmetric two-point function
$G^+(x,x^{\prime}).$ Equivalence principle suggest that the leading
singularity of $G^+(x,x^{\prime})$ should have a close
correspondence to singularity structure of the two-point function of
massless fields in Minkowski space [13]. In general the entire
singularity of $G^+(x,x^{\prime})$ may have a more complicated
structure. Usually one assumes that $G^+(x,x^{\prime})$ has a
singular structure represented by the Hadamard expansions. This
means that in a normal neighborhood of a point $x$ in 2d curved
space time, we can suggest logarithmic dependence (Hadamard Green
functions in 4d curved space times have singularities same as
$\sigma^{-1}$ and $\ln\sigma$ [1,2,15,16,17].) of the Green function
$G^+(x,x^{\prime})$ for a massless quantum scalar field $\chi$ as
\begin{equation}
G^{+}(x,x^{\prime})=V(x,x^{\prime})\ln\sigma(x,x^{\prime})+W(x,x^{\prime})\end{equation}
where $2\sigma(x,x^{\prime})=\sigma^a\sigma_a$ with
$\sigma_a\equiv\nabla_a\sigma$, is one-half square of the geodesic
distance between $x$ and $x^{\prime}.$ Nonsingular two point
functions  $V(x,x^{\prime}),$ and $W(x,x^{\prime})$  have the
following power series expansions \begin{equation}
V(x,x^{\prime})=\sum_{n=0}^{\infty}V_n(x,x^{\prime
})\sigma^n\end{equation} and \begin{equation}
W(x,x^{\prime})=\sum_{n=0}^{\infty}W_n(x,x^{\prime})\sigma^n\end{equation}
where $V(x,x')$ ($W(x,x')$) is state-independent (dependent) 2 point
functions. The Green function (28) satisfies the field equation (21)
with respect to both points $x$ and $x^{\prime}$ as\begin{equation}
\{\nabla_{c}\nabla^{c}-\frac{\nabla_{c}\nabla^{c}\Phi}{\Phi}\}G^{+}(x,x^{\prime})=
g^{-1/2}\delta^{2}(x-x^{\prime})\end{equation}
where $\delta^2(x-x^{\prime})$ is well known Dirac delta function in
2 dimensions. Applying (28), (29), (30) and (31), with
$x^{\prime}\neq x,$ the coefficients $V_n(x,x^{\prime})$ and
$W_n(x,x^{\prime})$ satisfies the following recursion relations.
\begin{equation}
2(n+1)^2V_{n+1}+2(n+1)\nabla_aV_{n+1}\sigma^a+(\nabla_{c}\nabla^{c}-
\frac{\nabla_{c}\nabla^{c}\Phi}{\Phi})V_{n}=0,\end{equation}
\begin{equation}(\nabla_{c}\nabla^{c}-\frac{\nabla_{c}\nabla^{c}\Phi}{\Phi})
W_n+2(n+1)\nabla_aW_{n+1}\sigma^a+2(n+1)^2W_{n+1}
$$$$+4(n+1)V_{n+1}+2\nabla_aV_{n+1}\sigma^a=0.\end{equation} Covariant
Taylor series expansion for symmetric  two point functions is
written as [15,16] (see also [17]) \begin{equation}
\Gamma(x,x^{\prime})=\Gamma(x)-\frac{1}{2}\nabla_a\Gamma(x)\sigma^a+\frac{1}{2}\Gamma_{ab}(x)\sigma^a\sigma^b$$$$
+\frac{1}{4}\{\frac{1}{6}\nabla_c\nabla_b\nabla_a\Gamma(x)-\nabla_c\Gamma_{ab}(x)\}
\sigma^a\sigma^b\sigma^c+O(\sigma^2)\end{equation} which for
$W(x,x')$ we obtain from coincidence limits \begin{equation}
W(x)=\lim_{x'\to x}W(x,x')=\lim_{x'\to
x}W_0(x,x')$$$$=W_0(x)=<\hat{\chi}^2>_{ren}=<\Phi^2\hat{\psi}^2>_{ren}.
\end{equation} The above renormalized expectation value is called vacuum state
of the quantum dilaton-matter field $\chi.$ Also one can obtain from
coincidence limits of the equations (32), (33), (34) and (35)
\begin{equation}
V_1(x)=\frac{1}{2}\left\{\frac{\nabla_{c}\nabla^{c}\Phi}{\Phi}V_0(x)-V_{0c}^c(x)\right\},\end{equation}
\begin{equation}
W_1(x)=V_{0c}^c(x)-\frac{W_{0c}^c(x)}{2}+\left(\frac{W_0(x)}{2}-V_0(x)
\right)\frac{\nabla_{c}\nabla^{c}\Phi}{\Phi}\end{equation} and
\begin{equation} W_{ab}(x)=W_{0ab}(x)+W_1(x)g_{ab}.\end{equation}
Applying (28) and (34) for $\Gamma(x,x')=V_0(x,x')$ the equation
(31) with $x^{\prime}\neq x$ leads to \begin{equation}
\{\nabla_{c}\nabla^{c}-\frac{\nabla_{c}\nabla^{c}\Phi}{\Phi}\}W(x,x^{\prime})=$$$$\frac{V_0(x)}{3}\{R_{ab}
\frac{\sigma^a\sigma^b}{\sigma}-\frac{1}{4}\nabla_aR_{bc}\frac{\sigma^a\sigma^b\sigma^c}{\sigma}
\}+O(\sigma)\end{equation} Inserting (34) for
$\Gamma(x,x')=W(x,x^{\prime}),$ the above equation reduces to the
following conditions. \begin{equation}
W^c_c(x)=\frac{\nabla_{c}\nabla^{c}\Phi}{\Phi}W_0(x)+\frac{V_0(x)}{3}R\end{equation}
and
\begin{equation}\nabla^b\left[3\widetilde{W}_{0ab}(x)+\frac{g_{ab}}{4}\left(V_0(x)R
-3\nabla_{c}\nabla^{c}
W_0(x)-\lambda(x)\right)\right]$$$$=R_{ae}\nabla^eW_0(x)\end{equation}
where\begin{equation}
\widetilde{W}_{0ab}(x)=W_{0ab}(x)-\frac{1}{2}g_{ab}W_{0c}^c(x)\end{equation}
and we used identities  \begin{equation} \nabla_{c}\nabla^{c}
\nabla^bW_0(x)=\nabla^b\nabla_{c}\nabla^{c}
W_0(x)+R^{ab}\nabla_aW_0(x),\end{equation}
\begin{equation}\nabla^b\nabla_a\nabla_b W_0(x)=\nabla_a\nabla_{c}\nabla^{c}
W_0(x)+R_{ab}\nabla^b W_0(x).\end{equation} We defined `effective
variable cosmological parameter` $\lambda(x)$ satisfying the
constraint condition \begin{equation}
R\nabla_aV_0(x)=\nabla_a\lambda(x)\end{equation} and also applied
\begin{equation}
V_{0a}^a(x)=V_0(x)\bigg(\frac{R}{6}+\frac{\nabla_{c}\nabla^{c}\Phi}{\Phi}\bigg),~~~V_1(x)=-\frac{V_0(x)R}{12}\end{equation}
\begin{equation}
W_{ab}(x)=\widetilde{W}_{0ab}(x)+g_{ab}\bigg(\frac{V_0(x)R}{6}+
\frac{W_0(x)}{2}\frac{\nabla_{c}\nabla^{c}\Phi}{\Phi}\bigg)\end{equation}
which are obtained from (36), (37), (38), (40). Now we subtract from
$G^+(x,x^{\prime})$ defined by (28), a local symmetric two point
function $G^+_L(x,x^{\prime})$ with the same short-distance
singularity of the Hadamard expansion. Then we make a renormaized
expectation value of stress tensor (25) as
\begin{equation} <\hat{T}^{ab}[\hat{\chi}]>_{ren}=\lim_{x^{\prime}\to x}
D^{ab}(x,x^{\prime})\{G^+(x,x^{\prime})-G^+_L(x,x^{\prime})\}\end{equation}
which by applying (28) can be rewritten as \begin{equation}
<\hat{T}^{ab}[\hat{\chi}]>_{ren}=\lim_{x^{\prime}\to x}
D^{ab}(x,x^{\prime})\{W(x,x^{\prime})\}.\end{equation} Explicit form
of the nonsingular stress tensor (49) is obtained by inserting (34)
[with $\Gamma(x,x')=W(x,x^{\prime})$], (47)  and taking its
coincidence limit as \begin{equation}
<\hat{T}_{ab}[\hat{\chi}]>_{ren}=\nabla_a\nabla_b
W_0(x)-2\widetilde{W}_{0ab}(x)-$$$$\frac{3}{2}(J_a\nabla_b+J_b\nabla_a)W_0(x)
+J_aJ_bW_0(x)+$$$$g_{ab}\{J_c\nabla^cW_0(x)-\frac{\nabla_{c}\nabla^{c}
W_0(x)}{2}-\frac{J^cJ_c}{2}W_0(x)\}\end{equation} where
$<\hat{T}^a_{a}[\hat{\chi}]>_{ren}=-J_c\nabla^cW_0(x).$ With same
calculation one can obtain for (24):  \begin{equation}
<g^{ab}\nabla_a\hat{\chi}\nabla_b\hat{\chi}>_{ren}=\lim_{x^{\prime}\to
x}D(x,x^{\prime})\{W(x,x^{\prime})\}=$$$$\frac{\nabla_{c}\nabla^{c}
W_0(x)}{2}-J_c\nabla^cW_0(x)+J_cJ^cW_0(x)-$$$$\frac{V_0(x)}{3}R+W_0(x)\frac
{\nabla_{c}\nabla^{c}\Phi}{\Phi}\end{equation} where we defined
\begin{equation}
D(x,x^{\prime})=g_a^{a^{\prime}}\nabla^a\nabla_{a^{\prime}}.\end{equation}
Applying (50) and identity (43) one can obtain
\begin{equation} \nabla^b\{<\hat{T}_{ab}[\hat{\chi}]>_{ren}
+2\widetilde{W}_{0ab}(x)+\frac{3}{2}(J_a\nabla_b+J_b\nabla_a)W_0(x)$$$$-J_aJ_bW_0(x)+
g_{ab}(\frac{1}{2}J_cJ^cW_0(x)-J_c\nabla^cW_0(x)$$$$-\frac{1}{2}\nabla_{c}\nabla^{c}
W_0(x))\}=R_{ae}\nabla^eW_0(x).\end{equation} Subtracting (41) from
(53) we obtain
\begin{equation}\nabla^a\Sigma_{ab}=0\end{equation} where $\Sigma_{ab}$
is general state independent divergence-less stress tensor relating
to $<\hat{T}_{ab}[\hat{\chi}]>_{ren}$ as \begin{equation}
<\hat{T}_{ab}[\hat{\chi}]>_{ren}=-\Sigma_{ab}+\widetilde{W}_{0ab}(x)+
J_aJ_bW_0(x)-$$$$\frac{3}{2}(J_a\nabla_b+J_b\nabla_a)W_0(x)+g_{ab}\{\frac{V_0(x)R}{4}-$$
$$\frac{\lambda(x)}{4}
-\frac{\nabla_{c}\nabla^{c}
W_0(x)}{4}-\frac{J_cJ^cW_0(x)}{2}+J_c\nabla^cW_0(x)\}\end{equation}
with
\begin{equation}
<\hat{T}^a_{a}[\hat{\chi}]>_{ren}=-\Sigma^a_a+\frac{V_0(x)R}{2}-
\frac{\lambda(x)}{2}$$$$-J_c\nabla^cW_0(x)-\frac{\nabla_{c}\nabla^{c}
W_0(x)}{2}.\end{equation}  The stress tensor $\Sigma_{ab}$ is really
geometric counterpart of the back reaction equation (16) in 2d
analogue and other terms in (55) denotes to matter dependent counter
part. This is subject which we seek to answer the question
 presented in last paragraph of the section 2 of the paper.
Inserting (51) into RHS of the equation (17) one can obtain
\begin{equation} \nabla_{c}\nabla^{c}
W_0(x)-2J_c\nabla^cW_0(x)+2\bigg(J_cJ^c+\frac{\nabla_{c}\nabla^{c}
\Phi}{\Phi}\bigg)W_0(x)$$$$=\left(\frac{2V_0(x)}{3}+\frac{\Phi^2}{4\pi
G}\right)R-\frac{\Phi\nabla_{c}\nabla^{c}\Phi}{2\pi
G}.\end{equation} Applying (56) and trace of the equation (16) we
obtain
\begin{equation}\Sigma^c_c =\frac{1}{4\pi
G}-\frac{\lambda(x)}{2}+\frac{V_0(x)R}{2}-\frac{\Phi\nabla_{c}\nabla^{c}\Phi}{4\pi
G}-\frac{\Phi^2J_cJ^c}{4\pi
G}$$$$-J_c\nabla^cW_0(x)-\frac{\nabla_{c}\nabla^{c}
W_0(x)}{2}.\end{equation} Applying the above relation and (55) the
backreaction equation (16) reduces to  \begin{equation}
\widetilde{\Sigma}_{ab}-\frac{1}{4\pi
G}\bigg\{\Phi\nabla_a\nabla_b\Phi-\frac{1}{2}g_{ab}\Phi\nabla_{c}\nabla^{c}\Phi\bigg\}=$$$$\widetilde{W}_{0ab}(x)+
W_0(x)\bigg[J_aJ_b-\frac{1}{2}g_{ab}J_cJ^c\bigg]$$$$
-\frac{3}{2}\bigg[(J_a\nabla_b+J_b\nabla_a)W_0(x)-g_{ab}J_c\nabla^cW_0(x)\bigg]\end{equation}
 where defined  \begin{equation}
\widetilde{\Sigma}_{ab}(x)=\Sigma_{ab}(x)-\frac{1}{2}g_{ab}\Sigma_{c}^c(x).\end{equation}
 Applying (51), (55), (58) and (59) the Bianchi identity (18) leads to the following constraint condition.
 \begin{equation}
  V_0(x)=\frac{3\Phi^2}{4\pi G}-\frac{3\Phi\nabla_{c}\nabla^{c}\Phi}{2\pi GR}+$$$$\frac{3\Phi^2J^b\nabla^a[\Phi^2J_cJ^c+5
  \Phi\nabla_{c}\nabla^{c}\Phi-2\Phi\nabla_a\nabla_b\Phi]
 }{8\pi GR(J_cJ^c)}.\end{equation}
Applying the above result one can obtain explicit form of the
cosmological parameter $\lambda(x)$ from (45) as
 \begin{equation}
\lambda(x)=\int R(x)\nabla_aV_0(x)dx^a+Constant .\end{equation} This
equation denotes to fluctuations of the variable cosmological
parameter $\lambda(x)$ satisfying to the wave equation
\begin{equation} \nabla_{c}\nabla^{c}\lambda(x)-\nabla^c\ln R\nabla_c\lambda=R\nabla_{c}\nabla^{c}
V_0(x).\end{equation} This wave equation is derived from constraint
condition (45) and its
RHS treats as geometrical source.\\
 However for a fixed 2d background metric $g_{ab}dx^adx^b,$
 we obtained 6 equations defined by (54), (57), (58), (59), (61) and (62)
which are not enough to determine seven quantities $W_0(x),
\lambda(x), V_0(x),\Phi(x),\Sigma_{ab}, \Sigma^c_c$ and
$W_{0ab}(x).$ Explicit form of all these quantities are depended to
form of the dilaton field $\Phi.$ What is its dynamical equation? In
particular spherically symmetric static space times with $\Phi(r)=r$
one can continue to solve the above equations and obtain the
foregoing dynamical fields but this is a bad restriction on our
procedure. For general dynamical 4d spherically symmetric curved
space times we should be have other management. Usually energy
conditions play important role on the physical sources. We study
energy conditions on 4d counter part of quantum matter stress tensor
given by (51), (55) and (56) in the following section.

\section{Energy conditions}
In general we consider time-like curves whose tangent 4-vector
$V^\mu=(V^a,0,0),$ with $V^\mu V_{\mu}=\beta>0,~a=0,1$ and
background metric signature $(-,+,+,+)$ which represents the radial
velocity vector of a family observer. In the latter case weak (WEC)
and strong (SEC) energy conditions leads to \begin{equation}
 WEC:~~~<\Phi^2\hat{T}_{ab}[\hat{\psi}]>_{ren}V^aV^b=\eta\geq0\end{equation}
 and \begin{equation}
 SEC:~~~<\Phi^2\hat{T}_{ab}[\hat{\psi}]>_{ren}V^aV^b-$$$$\frac{1}{2}\{<\Phi^2\hat{T}^a_{a}[\hat{\psi}]>_{ren}-
 <\Phi^2\partial_c\hat{\psi}\partial^c\hat{\psi}>_{ren}\}V^aV_a=\delta\geq0.\end{equation}
There is also a null energy condition (NEC) for radial null vector
field $N^\mu=(N^a,0,0)$ with $N^\mu N_\mu=0$ and $a=0,1$ as
\begin{equation}
NEC:~~~~~<\Phi^2\hat{T}_{ab}[\hat{\psi}]>_{ren}N^aN^b=\sigma\geq0.\end{equation}
Obviously, the above energy conditions emerge directly from the
geodesic structure of the spherically symmetric space time
(8).\\
Defining \begin{equation}
V^aJ_a=\alpha,~~~V^aV_a=\beta>0,~~~N^aJ_a=\gamma\end{equation} and
applying (51), (55), and (56) the energy conditions (64), (65) and
(66) leads to the following relations respectively.
\begin{equation}
WEC:~~~~~(W_{0ab}-\Sigma_{ab})V^aV^b+W_0(x)\bigg(\alpha^2-\frac{\beta}{2}
J^cJ_c\bigg)+$$$$(\beta J_c-3\alpha
V_c)\nabla^cW_0(x)+$$$$\frac{\beta}{4}\bigg(V_0(x)R-\lambda-\nabla_{c}\nabla^{c}
W_0(x)-2W^c_{0c}(x)\bigg)=\eta\end{equation} \begin{equation}
SEC:~~~~~~~\Sigma^c_c=\frac{2(\delta-\eta)}{\beta}-\frac{\lambda(x)}{2}+\frac{5V_0(x)}{6}R-\nabla_{c}\nabla^{c}
W_0(x)-$$$$\bigg(J_cJ^c+\frac{\nabla_{c}\nabla^{c}\Phi}{\Phi}\bigg)W_0(x)\end{equation}
and \begin{equation}
NEC:~~~~(W_{0ab}-\Sigma_{ab})N^aN^b+\gamma^2W_0(x)-$$$$3\gamma
N^c\nabla_cW_0(x)=\sigma.\end{equation} Applying (57) and (58), the
SEC given by (69) leads to the following wave equation.
\begin{equation} \nabla_{c}\nabla^{c}
\Phi^2-\bigg(J_cJ^c+\frac{R}{2}\bigg)\Phi^2=1+\frac{8\pi
G(\eta-\delta)}{\beta}\end{equation} where we used identity
$2\Phi\nabla_{c}\nabla^{c}\Phi+2\Phi^2
J_cJ^c=\nabla_{c}\nabla^{c}\Phi^2.$ This equation describes
evolutions of surface area of apparent horizon $S=4\pi \Phi^2$ of
the 4d spherically symmetric space time (7) propagating in 2d
induced space time $g_{ab}dx^adx^b.$ With (71), our strategy about
formulation of 2d analogue of the backreaction equation (1) and the
renormalized expectation value of the quantum matter-dilaton field
stress tensor operator is finished. It will be useful now we imply
apparent horizon property of the 4d spherically symmetric curved
space time (8) on our derived equations.
\section{Apparent Horizon}
Assuming $S=4\pi \Phi^2$ to be surface area of apparent horizon of
the spherically symmetric curved space time (8), one can obtain its
position by the null condition \begin{equation}
g^{ab}\nabla_aS\nabla_bS=0\end{equation} which by defining
$J_a=\nabla_a\ln\Phi$ leads to the condition \begin{equation}
J_aJ^a=0.\end{equation} In this case we can use $J_a=N_a$ as a
suitable null vector field in the NEC (66) for which $\gamma=0$ (see
(67)). In this case the NEC given by (70) leads to
\begin{equation} (W_{0ab}(x)-\Sigma_{ab})J^aJ^b=\sigma\geq0. \end{equation} Setting
$\sigma=0$ we can choose \begin{equation}
W_{0ab}(x)=\Sigma_{ab}(x)+\xi g_{ab}\end{equation} where $\xi$ is
arbitrary constant parameter. Using (73) and (75) the WEC (68) and
SEC (69) leads to respectively \begin{equation}
WEC:~~~~~~W_{0c}^c(x)=2\xi+\frac{V_0(x)R}{2}-\frac{\lambda(x)}{2}-\frac{\nabla_{c}\nabla^{c}
W_0(x)}{2}+$$$$\frac{2}{\beta}\{\alpha^2W_0(x)+(\beta J_c-3\alpha
V_c)\nabla^cW_0(x)-\eta\}\end{equation} and \begin{equation}
SEC:~~~~~~~\Sigma^c_c=\frac{2(\delta-\eta)}{\beta}-\frac{\lambda(x)}{2}+\frac{5V_0(x)}{6}R-$$$$\nabla_{c}\nabla^{c}
W_0(x)-\frac{\nabla_{c}\nabla^{c}\Phi}{\Phi}W_0(x).\end{equation}
 One of trivial solutions of the equation (45) is slow varying regime of the cosmological
 parameter $\lambda(x)$ for which we can exclude its derivatives as \begin{equation}
  \lambda(x)=\frac{4(\delta-\eta)}{\beta}\cong
 constant,~~~~~V_0(x)=\frac{1}{20\pi}.\end{equation} Under the latter assumptions the anomaly trace
 (5.6) become \begin{equation}
 \Sigma^c_c\cong\frac{R}{24\pi}-\omega\frac{\nabla_{c}\nabla^{c}\Phi}{\Phi}\end{equation}
in weak quantum field (WQF) limits as\begin{equation} W_0(x)\approx
constant=\omega>0
 \end{equation}
by excluding its derivatives. The anomaly trace (79) follows well
known one which is derived from zeta function regularization method
in 2d dilaton quantum field theory
[5,6,7,8,18,19,20,21,22,23,24,25,26,27,28,29,30] as
\begin{equation}
\Sigma^c_c(x)=\frac{1}{24\pi}\left\{R-\alpha\frac{\nabla_{c}\nabla^{c}\Phi}{\Phi}
+(\alpha-6)\frac{\nabla_c\Phi\nabla^c\Phi}{\Phi^2}\right\}
\end{equation} The arbitrary parameter $\alpha$ is the coefficient in
question [30]. $\alpha=-2$ proposed by R. Bousso and S. W. Hawking
[18] which turned out to be a mistake. $\alpha=4$ obtained by Kummar
et al. [7,19,20] for the same setup of the two-dimensional model as
was used by Bousso and Hawking. $\alpha=6$ obtained by Elizalde et
al. [21] and V. Mukhanov, A. Wipf and A. Zelnikov [5]. This result
turned out to be correct physically satisfying our statement about
apparent horizon induction on the anomaly. In other word (81)
reduces to (79) by setting \begin{equation}
\alpha=6,~~~~\omega=\frac{1}{4\pi}.\end{equation}
 In strong quantum field limits where we can not exclude
fluctuations of the field $W_0(x)$ and so its derivatives should be
considered in procedure one should be follow exact equations given
in the previous section. Generally, our procedure is useful to study
final state of quantum perturbed 4d spherically symmetric curved
space times. Asymptotically flat classical static metric solution of
the model (3) was obtained previously by Jains-Newman-Winicour (JNW)
[31,32]. As a future work one can use the presented formalism to
study physical effect of the obtained anomaly on the quantum
perturbed JNW metric solution. However we seek here slow varying
limits of the back reaction equation and obtain its vacuum sector
metric solution such as follows.
\section{Slow varying limits}
Setting
\begin{equation}
(\lambda(x),W_0(x),V_0(x))=(\Lambda,\omega,\mu)=constant\end{equation}
and
 \begin{equation} J_a=N_a,~~~J_aN^a=J^aJ_a=\gamma=0\end{equation}
 where we assumed $\Lambda$ to be the cosmological constant
 parameter the equations (56), (58), (59), (61), (68),
 (69), (70) and (71) reduce to the following forms respectively.
 \begin{equation}
 \bigg(\omega-\frac{\Phi^2}{4\pi G}\bigg)\frac{\nabla_c\nabla^c\Phi}{\Phi}
 =\bigg(\frac{2\mu}{3}+\frac{\Phi^2}{4\pi G}
 \bigg)\frac{R}{2}\end{equation}
  \begin{equation} \Sigma^c_c=\frac{1}{4\pi G}-\frac{\Lambda}{2}+\frac{\mu R}{2}-\frac{\Phi\nabla_c\nabla^c
  \Phi}{4\pi G}
 \end{equation}
  \begin{equation} \widetilde{W}_{0ab}(x)=
  \widetilde{\Sigma}_{ab}-\omega J_aJ_b-\frac{1}{4\pi G}\bigg\{\Phi\nabla_a\nabla_b\Phi-\frac{1}{2}g_{ab}\Phi\nabla_c\nabla^c
  \Phi
 \bigg\}
 \end{equation}
 \begin{equation}
  J^b\nabla^a(2\Phi\nabla_a\nabla_b-5\Phi\nabla_c\nabla^c
  \Phi)=0\end{equation}
\begin{equation}
\eta=(W_{0ab}(x)-\Sigma_{ab})V^aV^b+\alpha^2\omega+\frac{\beta}{2}[\mu
R-\Lambda-2W_{0c}^c(x)]\end{equation}
\begin{equation}
\Sigma^c_c=\frac{2(\delta-\eta)}{\beta}-\frac{\Lambda}{2}+\frac{5\mu
R}{6}-\frac{\omega\nabla_c\nabla^c
\Phi}{\Phi}\end{equation}
\begin{equation}
\sigma=(W_{0ab}(x)-\Sigma_{ab})J^aJ^b\end{equation}
\begin{equation}
\frac{\nabla_c\nabla^c
\Phi}{\Phi}-\frac{R}{4}=\frac{1}{\Phi^2}\bigg(\frac{1}{2}+\frac{4\pi
G(\eta-\delta)}{\beta}\bigg) .\end{equation} Applying (85), (86),
(90) and (92) one can results
\begin{equation}\eta=\delta\end{equation}
\begin{equation}
R=\frac{2}{\Phi^2}\bigg(\frac{\omega+\frac{\Phi^{2}}{4\pi
G}}{\frac{4\mu}{3}-\omega+\frac{\Phi^2}{4\pi G}}\bigg)\end{equation}
\begin{equation}\frac{\nabla_c\nabla^c
\Phi}{\Phi}=\nabla_cJ^c=\frac{1}{\Phi^2}\bigg(\frac{\frac{2\mu}{3}+\frac{\Phi^{2}}{4\pi
G}}{\frac{4\mu}{3}-\omega+\frac{\Phi^2}{4\pi G}}\bigg)\end{equation}
and
\begin{equation}\Sigma^{c}_{c}=\frac{\mu\bigg(\frac{5}{12\pi
G}-\frac{2\Lambda}{3}\bigg)+\omega\bigg(\frac{\Lambda}{2}-\frac{1}{4\pi
G}+\frac{\mu }{\Phi^2}\bigg)-\frac{\Lambda\Phi^{2}}{4\pi
G}}{\frac{4\mu}{3}-\omega+\frac{\Phi^2}{4\pi G}}.\end{equation}
Inserting (93), (94), (95), and (96) the equations defined by (87),
(88), (89) and (90) leads to the following forms respectively.
\begin{equation}\widetilde{W}_{0ab}(x)=\Sigma_{ab}+\bigg(\omega-\frac{\Phi^2}{4\pi
G}\bigg) J_aJ_b-\frac{\Phi^2\nabla_aJ_b}{4\pi
G}+$$$$\frac{g_{ab}}{2}\bigg[\Lambda\bigg(\frac{2\mu}{3}-\omega\bigg)+\frac{(\omega-3\mu)}{4\pi
G}-\frac{\mu\omega}{\Phi^2}+\bigg(\frac{1}{4\pi
G}+\frac{\Lambda}{2}\bigg)\frac{\Phi^2}{4\pi G}\bigg]\end{equation}
\begin{equation} J^b\nabla_c\nabla^c
J_b+3J^bJ^a\nabla_aJ_b=0,~~~J_a=\nabla_a\ln\Phi\end{equation}
\begin{equation}\eta=\frac{1}{24\pi
G}\bigg[\beta\bigg(\frac{3\omega-2\mu}{\frac{4\mu}{3}-\omega+\frac{\Phi^2}{4\pi
G}}\bigg)-6\Phi^2(\alpha^2+V^aV^b\nabla_aJ_b)\bigg]\end{equation}
and
\begin{equation}\sigma=\frac{\Phi^{2}J^b\nabla_c\nabla^c J_b}{12\pi G}.\end{equation} The equation
(97) determines only traceless part of the tensor $W_{0ab}(x).$ One
can obtain its trace part $W_{0c}^c(x)$ under the assumption
\begin{equation} \sigma=0\end{equation} for which the NEC is still satisfied.
In the latter case (98) and (100) is eliminated trivially leading to
the condition \begin{equation} J^b\nabla^c\nabla_c
J_b=0\end{equation} and if we insert (75), the equation (97) leads
to
\begin{equation}
W_{0c}^c(x)=2\xi-\Lambda\bigg(\frac{2\mu}{3}-\omega\bigg)-\frac{(\omega-3\mu)}{4\pi
G}+\frac{\mu\omega}{\Phi^2}$$$$-\bigg(\frac{1}{4\pi
G}+\frac{\Lambda}{2}\bigg)\frac{\Phi^2}{4\pi G}.\end{equation} We
are now in position to write explicit form of the Green function
(28) in terms of Hadamard series expansion. However one can rewrite
the non-linear dilaton wave equation (95) same as poisson equation
by defining a suitable dilaton field density $\rho(\Phi)$ such as
follows. \begin{equation} \nabla_bJ^b=-\frac{\rho(\Phi)}{4\pi
G}\end{equation} where we defined self-interaction dilaton field
density as
\begin{equation}\rho(\Phi)=\frac{4\pi
G}{\Phi^2}\left(\frac{\frac{2\mu}{3}+\frac{\Phi^2}{4\pi
G}}{\omega-\frac{4\mu}{3}-\frac{\Phi^2}{4\pi
G}}\right)\end{equation} where asymptotically flat Minkowski region
$\Phi\to\infty$ is free of dilaton field $\rho(\Phi)\to0.$ Also the
above density has a singularity at particular scale
$\Phi_s=\sqrt{4\pi G(\frac{4\mu}{3}-\omega)}$ where
$\omega\leq\frac{4\mu}{3}.$ We are now in position to solve the
metric backreaction equation.
\section{Backreaction metric solution}
It is useful to choose  conformaly flat frame in 2d space times for
which the 2d part of the background metric (8) is given by
\begin{equation} g_{ab}(x)dx^adx^b=e^{f(\Phi)}dudv\end{equation} where $(u,v)$ are suitable
null coordinates. In this case one can obtain corresponding 2d Ricci
scalar as \begin{equation} R=e^{-f(\Phi)}\nabla_c\nabla^c
f(\Phi)=e^{-f(\Phi)}f'(\Phi)\nabla_c\nabla^c\Phi\end{equation} where
over-prime $'$ denotes to differentiation with respect to the
dilaton field $\Phi$ and we used $J_cJ^c=0.$ Applying (94), (95) and
the above relation we obtain differential equation about the
conformal factor of the metric such as follows. \begin{equation}
e^{-f(s)}\frac{df(s)}{ds}=\frac{1}{s}\bigg(\frac{\omega+s}{\frac{2\mu}{3}+s}\bigg)\end{equation}
where we defined dimensionless apparent horizon surface area as
\begin{equation} s(\Phi)=\frac{\Phi^2}{4\pi G}.\end{equation} Integrating (108)
one can obtain \begin{equation}
e^{f(s)}=\frac{1}{(\frac{3\omega}{2\mu}-1)\ln\bigg(\frac{2\mu}{3}+s\bigg)-\frac{3\omega}{2\mu}\ln
s+C}\end{equation} where $C$ is a suitable integral constant which
should be fixed by using boundary condition of the space time. We
assume that the obtained metric (110) has apparent horizon from
point of view of a particular frame with coordinates
$(t,s,\theta,\varphi).$ In this case location of the apparent
horizon is obtained from the null-like condition of its hypersurface
$s_H=constant$ as $g^{ss}\partial_ss_H\partial_ss_H=0$ leading to
the condition $e^{-f(s_H)}=0.$ If we set $s_{H}=1$ (dimensionless
Plank length) and use $e^{-f(s_H)}=0$ then (110) leads to
\begin{equation}C=\bigg(1-\frac{3\omega}{2\mu}\bigg)\ln\bigg(1+\frac{2\mu}{3}\bigg).\end{equation}Using (111) the metric
solution (110) become
\begin{equation}
e^{f(s)}=\frac{1}{\bigg(\frac{3\omega}{2\mu}-1\bigg)\ln\bigg(\frac{\frac{2\mu}{3}+s}{\frac{2\mu}{3}+1}\bigg)-\frac{3\omega}{2\mu}\ln
s}.\end{equation}

\section{Concluding remarks}
In this article we used 2d analogue of the
 Einstein-massless scalar gravity to study 4d
 spherically symmetric quantum field theory. Hadamard
 renormalization prescription is used to obtain
 renormalized
 matter-dilaton stress tensor in the presence of variable
 cosmological parameter which has critical role to satisfy the stress tensor covariantly
  conservation condition. Singularity of the Hadamard Green function is assumed to be has logarithmic type
  same as the Green function in 2d Minkowski flat space
  time satisfying the general covariance condition. Our procedure  has an advantage with respect to other methods
   such as zeta function regularization:  Applying energy conditions (SEC, WEC, NEC)
    on the renormalized quantum matter dilaton field stress tensor
   we obtained  dynamical equations of the
  dilaton field $\Phi$, quantum vacuum state $W_0(x)$
  and variable cosmological parameter $\lambda(x)$ respectively.
  This is still an important problem in the Hadamard
  renormalization prescription used in general form of background metric in higher than 2 dimensions.
  In slow varying limits of quantum fields our obtained  anomaly trace
  satisfies
the well known one which is obtained from zeta function
regularization
  method. In the slow varying limits of the fields we solved the
  back reaction equation and obtained metric solution containing a
  horizon.

\bibliography{apssamp}
\vskip 0.5cm
\center
\textbf{REFERENCES}
\begin{description}
\item[1.] N. D. Birrell and P. C. W. Davies, \textit{Quantum Fields in Curved
space}, Cambridge University press, Cambridge, England, (1982).
 \item[2.] L. Parker and D. Toms \textit{Quantum Field Theory in Curved
space-time}, Cambridge University Press, Cambridge (2009).
\item[3.]
R. M. Wald, Phys. Rev. D17, 1477 (1978).
\item[4.] S. M. Christensen and S. A. Fulling, Phys. Rev. D15,
2088 (1977).
 \item[5.] V. Mukhanov, A. Wipe and A. Zelnikov,Phys. Lett B332, 283 (1994),
hep-th/9403018.
\item [6.] R. Balbinot and A. Fabbri, Phys. Rev. D59, 044031 (1999), hep-th/9807123.
\item[7.] W. Kummer, H. Liebl and D. V. Vassilevich, Mod. Phys. Lett. A12, 2683 (1997), hep-th/9707041.
\item [8.] R. Balbinot and A. Fabbri, Phys. Lett. B459, 112 (1999), gr-qc/9904034
\item [9.] P. Thomi, B. Isaak, and P. Hajicek, Phys. Rev. D30, 1168
(1984).
\item[10.] Shri Ram and M. K. Verma, ADVANCED RESEARCH in PHYSICS
AND ENGINEERING,  University of Cambridge, UK, February 20-22
(2010), Published by WSEAS press, ISSN: 1790-5117,
ISBN:978-960-474-163-2, Pages:23-28.
\item[11.] Antonio Alfono-Faus, Recent Researches in Artificial
Intelligence, Knowledge Engineering and Data Bases, University of
Cambridge, UK, February 20-22 (2011), Published by WSEAS press,
ISBN:978-960-474-273-8, Pages:249-254.
\item[12.] Andrew Waloott Bechwith, Advances in Applied and Pure
Mathematics, Gdansk University of Technology, Poland, May 15-17
(2014),Published by WSEAS press, ISBN:978-960-474-380-3,
Pages:313-322.

\item[13.] R. Haag. H. Narnhofer and U. Stein, Commun. Math. Phys.
94, 219, (1984).
\item[14.] M. Gasperini, \textit{Elements of String Cosmology}, Cambridge University press (2007).
\item[15.] M. R. Brown, J. Math. Phys.25(1), 136 (1984).
\item[16.]  D. Bernard and A. Folacci, Phys. Rev. D34, 2286 (1986).
\item[17.] H. Ghafarnejad and H. Salehi, Phys. Rev. D 56, 4633,
(1997); 57, 5311 (E) (1998).
\item[18.] R. Bousso and S. W. Hawking, Phys. Rev.
D56, 7788 (1997).
\item[19.] W. Kummer, H. Liebl and D. V. Vassilevich, Phys. Rev.
D58, 108501 (1998), hep-th/9801122.
\item[20.] S. Ichinose, Phys. Rev. D57, 6224, (1998), hep-th/9707025.
\item[21.]E. Elizalde, S. Naftulim, S. O. Odintsov, Phys. Rev. D49,
2852 (1994), hep-th/9308020.
\item[22.] S. Nojiri and S. D. Odintsov, Mod. Phys. Lett. A12, 2083
(1997), hep-th/9706009.
\item[23.] S. Nojiri and S. D. Odintsov, Phys. Rev. D57, 2363
(1998), hep-th/9706143.
\item[24.] S. Nojiri and S. D. Odintsov, Phys. Lett B416, 85 (1998),
hep-th/9708139.
\item[25.] S. Nojiri and S. D. Odintsov, Phys. Lett. B426, 29
(1998), hep-th/9801052.
\item[26.] S. Nojiri and S. D. Odintsov, Phys. Rev. D57, 4847
(1998), hep-th/9801180.
\item[27.] S. J. Gates Jr, T. Kadoyoshi and S. D. Odintsov, Phys.
Rev. D58, 084026 (1998), hep-th/9802139.
\item[28.] P. Van Nieuwenhuizen, S. Nojiri and S. D. Odintsov, Phys.
Rev. D60, 084014 (1999), hep-th/9901119.
\item[29.] S. Nojiri and S. D. Odintsov, Int. J. Mod. Phys. A16,
1015 (2001), hep-th/0009202.
\item[30.] J. S. Dowker, Class. Quantum Grav 15, 1881 (1998), hep-th/9802029.
\item [31.] A. I. Janis, E. T. Newman and J. Winicour, Phys. Rev. Lett. 20, 878, (1968).
\item[32.] K. S. Virbhadra, D. Narasimha and S. M. Chitre, Astron. Astrophys. 337, 1-8 (1998).
\end{description}
\end{document}